\documentclass[epj]{svjour}
\usepackage{graphics}
\begin{document}
\title{Bulk rheology and microrheology of active fluids}
\author{G. Foffano\inst{1}, J.~S. Lintuvuori\inst{1}, A.~N. Morozov\inst{1}, K. Stratford\inst{2}, M.~E. Cates\inst{1}, D. Marenduzzo\inst{1}%
}                     
\offprints{}          
\institute{SUPA, School of Physics and Astronomy, University of Edinburgh, Mayfield Road, Edinburgh EH9 3JZ, UK \and EPCC, School of Physics and Astronomy, University of Edinburgh, Mayfield Road, Edinburgh, EH9 3JZ, UK}

\date{Received: date / Revised version: date}
%
\abstract{
We simulate macroscopic shear experiments in active nematics and compare them with microrheology simulations where a spherical probe particle is dragged through an active fluid. In both cases we define an effective viscosity: in the case of bulk shear simulations this is the ratio between shear stress and shear rate, whereas in the microrheology case it involves the ratio between the friction coefficient and the particle size. We show that this effective viscosity, rather than being solely a property of the active fluid, is affected by the way chosen to measure it, and strongly depends on details such as the anchoring conditions at the probe surface and on both the system size and the size of the probe particle. 
} 

\authorrunning{G. Foffano et al.}
\titlerunning{Bulk rheology and microrheology of active fluids}
\maketitle
\section{Introduction}
\label{intro}
Active fluids are suspensions of particles that absorb energy from their surroundings (or from an internal reservoir) in order to do work. They represent a novel type of out-of-equilibrium system and they are found predominantly, although not solely, in biological contexts~\cite{sriram_review}. Active particles exert forces on the fluid in which they are embedded and, to lowest order, they can be modelled as a force dipole. Extensile active particles exert forces that are directed outward, along their main axis, while contractile particles exert forces directed inward. Examples of the former are suspensions of bacteria like \textit{E.coli}, while dispersions of algae, such as {\it Chlamidomonas}, and the actomyosin solution which constitutes the cytoskeleton of eukariotic cells, are contractile fluids.

Due to the continuous  stirring exerted by their constituent particles, active fluids show a wide range of novel properties, the most striking of which is probably the appearence of a {\em spontaneous flow} phase: for high enough dipolar forces or active particle density, these fluids can flow spontaneously in steady state, in the absence of any external force~\cite{spontaneousflow_I,spontaneousflow_II,spontaneousflow_III,spontaneousflowvoituriez,goldstein}. 

The bulk rheological properties of active fluids are also highly non-standard. In particular, theory predicted~\cite{Ramaswamy_PRL_2004,Liverpool_Marchetti_PRL_2006}, and both simulations~\cite{fieldingPRL} and experiments~\cite{rafai_PRL,Aranson} confirmed, that active nematics increase their viscosity with activity if contractile and decrease it if extensile. On the other hand, microrheology experiments, where the motion of micron-sized probe particles is studied~\cite{waigh}, allow one to investigate {\it local}, rather than global, properties which are important in biophysical contexts. (For instance, cells are often subjected to localised cues rather than to bulk external body forces~\cite{bray}.) These experiments have also shown strong violations of the fluctuation-dissipation theorem~\cite{lubensky,mizuno}, further proving that the physics of active fluids has little in common with that of passive, near-equilibrium systems.

Here we present a systematic numerical study of the rheological properties of active fluids, which we investigate both globally, through shear experiments (in a quasi-one dimensional geometry, as used in~\cite{fieldingPRL}), and locally, by applying a constant force on a spherical colloid embedded in the fluid.
The resulting comparison is new, and important, as it allows us to probe to what extent an effective viscosity, $\eta_{\rm eff}$, exists for active fluids: for this concept to be a useful one, $\eta_{\rm eff}$ should not depend much on the way in which the behaviour of the active fluid is probed. It hence should have the same or a similar value in both bulk rheology and microrheology experiments.
For bulk rheology simulations the effective viscosity is defined as the ratio between the steady state shear stress and the applied shear rate. For microrheology, if we measure the steady state velocity $v$ reached by a spherical particle subjected to a pulling force $F$, the effective viscosity can be defined as $\eta_{\rm eff}=\xi_0/6\pi R$, with $\xi_0 = F/v$ the friction coefficient. 

Our bulk rheology simulations show that in sheared nematics $\eta_{\rm eff}$ differs for different anchoring conditions at the container walls, as expected in view of the anisotropic nature of the fluid. However, the effective viscosity also depends on system size, in a way which it is possible to predict very accurately from an analysis of the underlying equations of motion.
With our microrheology simulations, instead, we find that, when a spherical particle is dragged through an active fluid, the drag force does not depend linearly on the particle radius, $R$, in strong violation of Stokes' law. We first present our results in the case of planar anchoring at the probe surface and we explain this effect in terms of the active stress distribution and the director-field deformation determined by the anchoring condition on the particle surface. To test this reasoning we consider also the case of normal anchoring and the extreme case in which no anchoring condition is imposed, so that the far field order is kept also in the proximity of the particle.

While we performed most microrheology simulations with periodic boundary conditions, we also checked the role of boundary walls. To this end, we repeated the microrheology experiments relative to planar anchoring at the probe surface, when the sample was bounded by walls along one direction, and found good qualitative agreement with the case of periodic boundary conditions.
Finally, in order to connect more easily to future experiments, we modelled also the case in which the particle disrupts the active fluid ordering, by imposing zero order parameter on its surface. We did this because the particles used in microrheology experiments are usually micron-sized, so that the size of the microrheological probe may be in practice similar to that of the active particles (e.g. suspensions of bacteria, such as {\it E.coli}). In this case one may envisage that, rather than creating homogeneous/normal anchoring, the probe particle may simply locally create a more disordered region. 

Our paper is structured as follows. In Section~\ref{sec:1} we present our model. In Section~\ref{sec:2} we discuss our results, first for bulk rheology (Section~\ref{sec:2.1}) and then for simulations on microrheology (Section~\ref{sec:2.2}). In Section~\ref{sec:3} we present further discussion on our results and draw conclusions. 
Some elements of the microrheology work were reported briefly in~\cite{PRL}, which mainly focuses however on a regime of high activity/large probe size that is not addressed here.

\section{Model and methods}
\label{sec:1}
In the continuum limit, active fluids can be described within a phenomenological hydrodynamic model, which is a generalisation of the equations of motion describing liquid crystal hydrodynamics. The additional terms which we introduce account for the effect of forces exerted by the active particles on the fluid. 
Liquid crystals are often described in terms of the director $\vec{n}$, a head-tail symmetric vector describing the local average orientation of the rodlike particles which typically constitute them. A more general characterisation can be achieved by defining a tensorial order parameter $\mathsf{Q}$, which can account for the average magnitude of orientational order, on top of its direction. 
The magnitude of order, $S$, and the director field, $\vec{n}$, are then associated to respectively the largest eigenvalue of $\mathsf{Q}$ and its associated eigenvector. 

In this work, we start from a generic Landau - de Gennes free energy density $\mathcal{F}=\mathcal{F}_b+\mathcal{F}_{el}+\mathcal{F}_s$, which describes the equilibrium physics of liquid crystals and consists of a bulk, of an elastic, and of a surface contribution. The bulk contribution, in the nematic case, takes the following form~\cite{deGennes}:
\begin{equation}\label{eq:fed_bulk}
\mathcal{F}_b = \frac{A_0}{2}\left(1-\frac{\gamma}{3}\right)Q_{\alpha \beta}^2-\frac{A_0 \gamma}{3}Q_{\alpha \beta}Q_{\beta \gamma}Q_{\gamma \alpha} + \frac{A_0 \gamma}{4}(Q_{\alpha \beta}^2)^2, 
\end{equation}
where $A_0$ is a constant, with dimension of free energy density, while the parameter $\gamma$ controls the magnitude of order -- this is related physically to either the temperature (in thermotropic liquid crystals), or the concentration (in lyotropic liquid crystals). The transition from the isotropic to the nematic phase takes place at $\gamma = 2.7$; here we work at $\gamma = 3.0$, which leads to an ordered phase. The elastic free energy density is:
\begin{equation}\label{eq:fed_el}
\mathcal{F}_{el}=\frac{K}{2}(\partial_{\beta}Q_{\alpha \beta})^2,
\end{equation}
where $K$ is an elastic constant (we fixed it to $K=0.05$ in simulation units (s.u.) in what follows, see below for a mapping between s.u. and physical units).
The one elastic constant approximation which we choose here does not modify the physics we are interested in qualitatively (for an example of the effect of using different splay and bend elastic constant in active fluids, without colloids, see e.g.~\cite{jenssoftmatter}).
Note that in Eqs.~(\ref{eq:fed_bulk}) and (\ref{eq:fed_el}), and below, Greek indices are used to denote Cartesian coordinates, and that summation over repeated indices is implied. 

The surface free-energy density term that describes the anchoring of nematogens to a solid surface (either a wall or a colloidal surface) is:
\begin{equation}\label{eq:fed_surf}
\mathcal{F}_s=\frac{1}{2}W(Q_{\alpha \beta}-Q_{\alpha \beta}^0)^2, 
\end{equation}
where $Q_{\alpha \beta}^0 = S_0(n_{\alpha}^0 n_{\beta}^0-\delta_{\alpha \beta}/3)$, $S_0$ quantifies the preferred degree of order, and $\vec{n}^0$ describes the director orientation on the surface~\cite{Nobili}.  Different types of anchoring are implemented in our code by fixing $Q^0$ accordingly (see~\cite{Juho_JMC,Juho_PRL}, e.g. tangential anchoring is enforced by projecting the order parameter on the local colloidal surface~\cite{Fournier}), whereas $W$ is the parameter controlling the anchoring strength. The boundary conditions for the order parameter on the colloidal surface are given by~\cite{juho_dimers,zumer_colloid}:
\begin{equation}
\nu_{\gamma}
\frac{\partial \mathcal{F}_s}{\partial \partial_{\gamma} Q_{\alpha\beta}}
+\frac{\partial\mathcal{F}_s }{\partial Q_{\alpha\beta}}=0
\end{equation}
where $\nu_{\gamma}$ is the local normal to the colloid surface.
 
The hydrodynamic equation for the evolution of the order parameter is:
\begin{equation}\label{eq:OP}
(\partial_t+\vec{u}\cdot\vec{\nabla})\mathsf{Q} - \mathsf{S}(\nabla \vec{u},\mathsf{Q})=\Gamma\mathsf{H},
\end{equation}
where $\Gamma$ is a collective rotational diffusion coefficient related to the rotational viscosity of the liquid crystal~\cite{spontaneousflow_I,spontaneousflow_II,spontaneousflow_III}.
The second term of eq.~(\ref{eq:OP}) describes the advection of the order parameter; while $\mathsf{S}(\nabla \vec{u},\mathsf{Q})$ describes the possibility for $\mathsf{Q}$ to be rotated and stretched by the fluid and takes the form~\cite{Beris-Edwards}:
\begin{eqnarray}\nonumber
&&\mathsf{S}(\nabla \vec{u},\mathsf{Q})=(\xi \mathsf{D}+\omega)(\mathsf{Q}+\mathsf{I}/3)+\\
&&+(\mathsf{Q}+\mathsf{I}/3)(\xi\mathsf{D}-\omega)-2\xi (\mathsf{Q}+\mathsf{I}/3)\mbox{Tr}(\mathsf{Q}\nabla \vec{u}).
\label{S}
\end{eqnarray}
In Eq.\ref{S} $\mbox{Tr}$ denotes the tensorial trace, while $\mathsf{D}=(\nabla \vec{u}+ \nabla \vec{u}^T)/2$ and $\omega=(\nabla \vec{u}-\nabla \vec{u}^T)/2$ are respectively the symmetric and the antisymmetric parts of the velocity gradient tensor.
The right-hand side of eq.~(\ref{eq:OP}) describes the relaxation of the order parameter towards an equilibrium configuration. The tensor $\mathsf{H}$ ({\it molecular field}) represents the symmetric and traceless part of the functional derivative of the free energy $F=\int d^3x \mathcal{F}(\vec{x})$ with respect to the order parameter $Q_{\alpha \beta}$:
\begin{equation}
H_{\alpha \beta}=-\frac{\delta F}{\delta Q_{\alpha \beta}}+\frac{\delta_{\alpha \beta}}{3}\mbox{Tr}\frac{\delta F}{\delta Q_{\alpha \beta}}.
\end{equation} 
The evolution of fluid momentum for an incompressible fluid ($\partial_\alpha u_\beta=0$) is described by a modified Navier-Stokes equation where new stress terms are introduced:
\begin{equation}\label{eq:NS}
\rho(\partial_t+u_\beta \partial_\beta)u_\alpha = \eta \partial_\beta(\partial_\beta u_\alpha + \partial_\alpha u_\beta)+\partial_\beta (\Pi_{\alpha \beta}^p+\Pi_{\alpha \beta}^a).
\end{equation}
Here $\eta$ is the bare viscosity of the Newtonian fluid (set to $0.6$ in s.u.), while $\Pi_{\alpha \beta}^p$ is the stress tensor of a passive liquid crystal, whose detailed expression is~\cite{Beris-Edwards}:
\begin{eqnarray}\label{eq:stress_passive}\nonumber
&&\Pi_{\alpha \beta}^{p} = -P_0\delta_{\alpha \beta}+2\xi \left( Q_{\alpha \beta}+\frac{1}{3}\delta_{\alpha \beta}\right)Q_{\gamma \epsilon}H_{\gamma \epsilon}\\
\nonumber
&&-\xi H_{\alpha \gamma} \left( Q_{\gamma \beta} + \frac{1}{3}\delta_{\gamma \beta}\right) - \xi \left(Q_{\alpha \gamma}+\frac{1}{3}\delta_{\alpha \gamma} \right)H_{\gamma \beta}\\
&&-\partial_\alpha Q_{\gamma \nu} \frac{\delta \mathcal{F}}{\delta \partial_\beta Q_{\gamma \nu}} + Q_{\alpha \gamma}H_{\gamma \beta} - H_{\alpha \gamma}Q_{\gamma \beta},
\end{eqnarray}
where $P_0$ is the isotropic pressure.
The parameter $\xi$ determines whether the liquid crystal is flow tumbling (which happens for $\xi<0.6$) or flow aligning ($\xi>0.6$). In the absence of activity but with flow present, particles in flow aligning liquid crystals assume a fixed angle with respect to the flow direction (Leslie angle), while in flow tumbling they continuously change their orientation, chaotically. To avoid further complexity arising from this, we fixed $\xi=0.7$ (s.u.).

Hydrodynamic equations for active fluids are obtained through an additional stress term in eq.~\ref{eq:NS}, which takes into account the effect of active forces, and is given by:
\begin{equation}\label{eq:stress_active}
\Pi_{\alpha \beta}^a=-\zeta \left(Q_{\alpha \beta}+\frac{1}{3}\delta_{\alpha \beta}\right),
\end{equation}
where $\zeta$, the activity coefficient, is a constant related to the size of active particles, to their density and to the intensity of the forces they exert on the ambient fluid. The parameter $\zeta$ is negative for contractile and positive for extensile liquid crystals~\cite{sriram_review}. 

We solve these equations through a hybrid Lattice-Boltzmann method (described in~\cite{Juho_JMC,Juho_PRL}), where eq.~\ref{eq:OP} is solved through a finite-difference algorithm, while the Lattice-Boltzmann method is used for momentum transport (Navier-Stokes equation and continuity). Within this framework, a spherical probe colloid is simulated through the method of bounce-back on links, which allows us to implement the no-slip boundary condition at the colloid surface~\cite{Stratford_BBL}. Different stress terms appearing in eq.~\ref{eq:NS}, have to be integrated over the colloid surface to determine the force and the torque acting on the colloid due to the surrounding fluid~\cite{Juho_JMC,Juho_PRL,juho_dimers}.

Above and in what follows, parameters and results are presented in simulation units. To convert them into physical ones, relative to a contractile actomyosin solution, we can consider values holding for typical intracellular actin gels~(\cite{mogilner}): the elastic constant $K$ is 1.25 nN, and the rotational viscosity is 10 kPa/s.
In this way (see~\cite{spontaneousflow_I,spontaneousflow_II,spontaneousflow_III} for similar mappings) one simulation unit for forces corresponds to 25nN, the lattice unit for length, $\Delta x$, is 0.5 $\mu$m, while the time unit corresponds to 10 ms~\cite{noteactomyosin}. Note that the mapping to a different active fluid, such as a bacterial suspension, would be significantly different~\cite{spontaneousflow_I,spontaneousflow_II,spontaneousflow_III}. Finally, the anchoring strength $W$, when different from zero, is set to $K/\Delta x$.

Note that the above equations assume nematic order, in which individual active particles, which generally are polar, are equally likely to point in any direction along $\vec{n}$. For polar phases different equations are needed (see e.g.~\cite{Giomi_Marchetti_2012}). Eq.~(\ref{eq:stress_active}) however holds at first order for both polar and apolar phases (see~\cite{sriram_review,Ramaswamy_PRL_2004}), and we therefore expect rheological properties not to strongly depend on polarity. We will thus restrict ourselves to the apolar case in this work.

We also point out that, as we employ a continuous model for active liquid crystals, our results on microrheology are reliable only as far as we address colloids with a size much larger than the size of the active nematogens, e.g. the length of a bacterium, or the mesh size of an actomyosin solution. Typical probes  in microrheology are in the 1-10 $\mu$m range. Therefore in order to describe bacterial or algal microrheology within the continuum active gel theory, one may need to use rather large probes. However this restriction is lifted in the case of contractile actomyosin gels. 

\section{Results}
\label{sec:2}
\subsection{Macrorheology}
\label{sec:2.1}

\begin{figure*}
\resizebox{0.9\textwidth}{!}{%
  \includegraphics{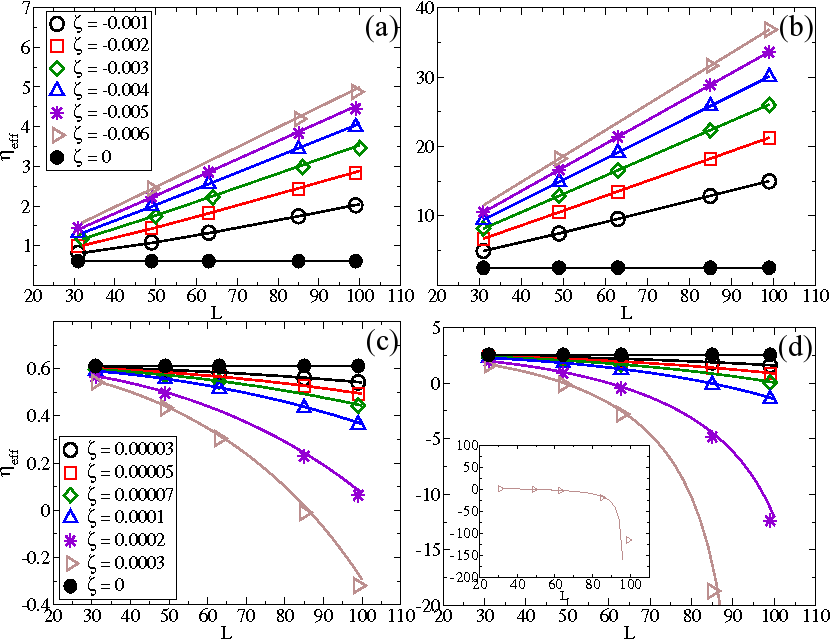}
}
\caption{Dependence of $\eta_{\rm eff}$ on system size $L$ in the case of contractile active nematics with planar (a) and normal (b) anchoring, and an extensile active fluid with planar (c) and normal (d) anchoring at the walls. Legend on the left refers to both graphs in the same line. The inset in (d) shows the behaviour for $\zeta=0.0003$ more in detail. Points here refer to numerical results, while solid lines are theoretical predictions obtained through Eq.~\ref{eq:theory}.}
\label{fig:1}       
\end{figure*}

Bulk rheological properties of active fluids have been investigated, theoretically~\cite{Ramaswamy_PRL_2004,Liverpool_Marchetti_PRL_2006}, with simulations~\cite{fieldingPRL,smfpre}, and through experiments~\cite{rafai_PRL,Aranson}.

Given the importance of boundary conditions in determining the behaviour of liquid crystalline and of active fluids (as found e.g. in~\cite{spontaneousflow_I,spontaneousflow_III,fieldingPRL}), we address here the question of the role played by the anchoring condition on the walls and by the system size. With this aim in mind, we consider two different cases: planar anchoring, where active nematogens are oriented along the walls, and normal anchoring.

As anticipated in the Introduction, an effective viscosity $\eta_{\rm eff}$ can be defined as the ratio between the measured stress and the imposed average shear rate, and we measure it for different values of $\zeta$.
We highlight that here we are interested in the bulk microrheology of active fluids in their quiescent (passive) phase, i.e. when they are not spontaneously flowing. We focus here for simplicity on the case of a quasi-one-dimensional system, where there is translational invariance in the vorticity and flow direction, and all quantities only vary along the flow gradient direction, which is $z$ in our notation. This case is already sufficient to describe qualitatively the basic features of the spontaneous flow transition in active fluids~\cite{spontaneousflow_I,spontaneousflow_II,spontaneousflow_III,fieldingPRL}. Spontaneous flow can be avoided by applying a small enough stress. 

Fig.~\ref{fig:1} shows that $\eta_{\rm eff}$ in the nematic phase in our geometry behaves consistently with previous theoretical expectations~\cite{Ramaswamy_PRL_2004,Liverpool_Marchetti_PRL_2006}: it is higher with respect to the passive isotropic viscosity in the case of contractile active fluids and smaller in the extensile case. Moreover, by analysing its dependence on the system size $L$, which we define as the distance between the two shearing planes, we find that $\eta_{\rm eff}$ increases with $L$ in contractile active fluids and decreases with $L$ in extensile ones. 
As can be seen from Fig.~\ref{fig:1}, the anchoring of the active particle orientation at the walls plays quantitatively an important role: for instance the difference with the passive viscosity is enhanced for normal anchoring. 

An analytical prediction for the apparent viscosity $\eta_{\rm eff}$ at low shear rates is obtained by linearising the equations of motion, Eqs.~\ref{eq:OP},\ref{eq:NS}. Assuming that the flow is in the $y$-direction, we obtain
\begin{eqnarray}
&&\Gamma K \delta Q_{yz}''(z) + \frac{\alpha}{2} u_y'(z) =0, \\
&&\eta u_y''(z) - \zeta \delta Q_{yz}'(z) - K x \delta Q_{yz}'''(z) = 0,
\end{eqnarray}
where $\delta Q_{yz}(z)$ and $u_y(z)$ are the deviations of the order-parameter tensor and the velocity from their rest values due to the applied shear; prime denotes the derivative with respect to $z$. Here,
\begin{equation}
\alpha = \xi-1 \qquad x=\frac{2\xi+S(\xi-3)}{3},
\end{equation}
for the planar anchoring, and 
\begin{equation}
\alpha = \xi+1 \qquad x=\frac{2\xi+S(\xi+3)}{3},
\end{equation}
for the normal anchoring. The linear reponse of the system to the applied shear, satisfying the boundary conditions $u_y(0)=-v_0$,  $u_y(L)=v_0$,  and $\delta Q_{yz}(0)=\delta Q_{yz}(L)=0$, is given by
\begin{eqnarray}
&&u_y(z) = v_0 \frac{\sinh{\lambda z}}{\sinh{\frac{\lambda L}{2}}}, \\
&&\delta Q_{yz}(z) =\frac{\alpha\, v_0}{2\Gamma K \lambda  \tanh{\frac{\lambda L}{2}}} \left( 1-  \frac{\cosh{\lambda z}}{\cosh{\frac{\lambda L}{2}}} \right),
\end{eqnarray}
where
\begin{equation}
\lambda=\sqrt{\frac{\zeta x}{2\eta \Gamma K+K x^2}}.
\end{equation}
The apparent viscosity is calculated by dividing the total shear stress by the shear rate $2 v_0/L$ which yields
\begin{equation}\label{eq:theory}
\eta_{\rm eff}=  \frac{L \lambda}{2 \tanh{\frac{\lambda L}{2}}} \left[ \eta+\frac{x^2}{2\Gamma} \right].
\end{equation}

In Fig.~\ref{fig:1} we compare this analytical prediction (solid lines) with our lattice Boltzmann data (symbols). The results show a very good agreement between the viscosity in the linear regime obtained in our simulations with the predictions in Eq.~\ref{eq:theory}, except for the case of an extensile fluid with $L=100$ and $\zeta=0.0003$, where spontaneous flow sets up in the numerical solution at $\dot{\gamma}=2 \cdot 10^{-7}$, which introduces non-linear effects invalidating our linearised analytical treatment. 

A key feature of our results is that the effective viscosity for the active (but not passive) system strongly depends on system size, so that this quantity would be ill defined in a system where the walls are infinitely far apart, clearly at odds with what is expected of a ``well-behaved'' macroscopic viscosity of a fluid sample. In other words, the viscosity of an active fluid cannot be operationally defined as the ratio between shear stress and shear rate, as routinely done in experiments probing the bulk rheology of complex fluids. Notably, as Fig.~\ref{fig:1} shows, this is despite the existence of a linear regime in the flow curve, which does appear and can be fitted with our analytical formulas.

\subsection{Microrheology}
\label{sec:2.2}
\begin{figure*}
\resizebox{0.9\textwidth}{!}{%
  \includegraphics{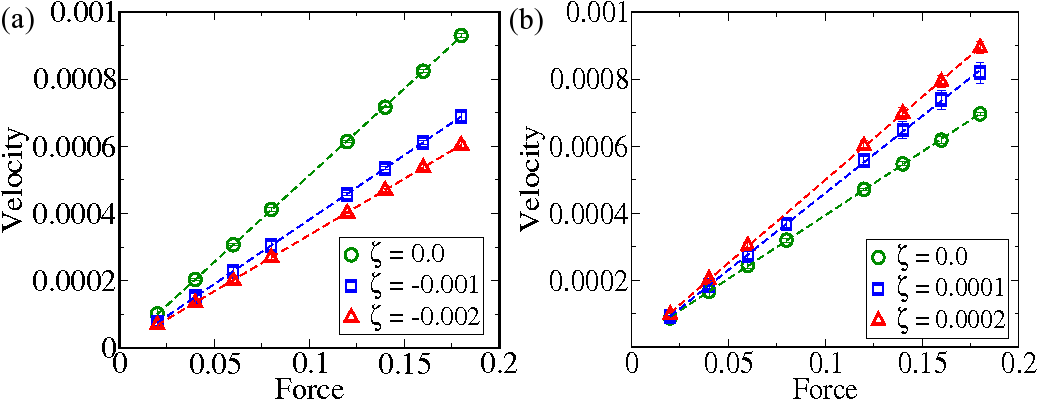}
}
\caption{Dependence of the particle velocity, measured in steady state, on the external force for a particle of fixed radius $R=11.3$, and different $\zeta$ (see legend). (a) refers to a contractile and (b) to an extensile active fluid. We used (periodic) cubic simulation box of volume $V=128^3$. }
\label{fig:2}       
\end{figure*}

We now turn to the {\em micro}rheology of active fluids. To study this, we performed simulations in which a spherical particle is dragged by a constant force through an active nematic, considering the cases in which the dragging direction is either parallel or orthogonal to the average direction of the nematogens in the bulk, which we will refer to as the far field director. 
These studies simulate the simplest class of microrheological experiments: for passive fluids in the linear regime, where the drag force on the particle depends linearly on its velocity, one would expect to see Stokes' law at work. This prescribes that the drag coefficient should depend linearly on particle size and on the fluid viscosity. Stokes' law is indeed typically used in experiments to infer mean properties of the fluid, in particular its viscosity.

Here we first monitor the colloid velocity in steady state, $v$, as a function of the magnitude of the applied force, $F$. From Fig.~\ref{fig:2} it can be seen that these two quantities are linearly related, thereby allowing us to define the drag on the particle as $\xi_0=F/v$. We then define an effective viscosity via Stokes' law as $\eta_{\rm eff}=\xi_0/6\pi R $.

We performed most of our simulations with periodic boundary conditions, for which corrections taking into account hydrodynamic interactions between different images of the same particles are needed. These are calculated in the case of a simple fluid in~\cite{Hasimoto}, while to our knowledge no similar calculations were done for liquid crystals. However, we found that~\cite{Hasimoto} seems to work quite well also in this case, since corrected data for passive nematics do not show significant dependence of $\eta_{\rm eff}$ on $R$. We then applied the same corrections in the case of active fluids.  

Just as we considered different system sizes, $L$, in the macrorheology section, here we study the behaviour for different particle sizes $R$.

We chose values of $\zeta$ high enough for activity to affect the system behaviour, but small enough not to see any spontaneous bulk flow effects. 
In particular, we considered $\zeta=-0.001$ and $-0.002$ for contractile nematics and $\zeta=0.0001$ and $0.0002$ for extensile fluids~\cite{activity_oom}, in the case of planar anchoring. In all other cases we chose a single value for $\zeta$ ($\zeta = -0.001$ for contractile and $\zeta=0.0001$ for extensile nematics), as we wish to focus on the effect of anchoring on $\eta_{\rm eff}$. 

\begin{figure*}
\resizebox{0.9\textwidth}{!}{%
  \includegraphics{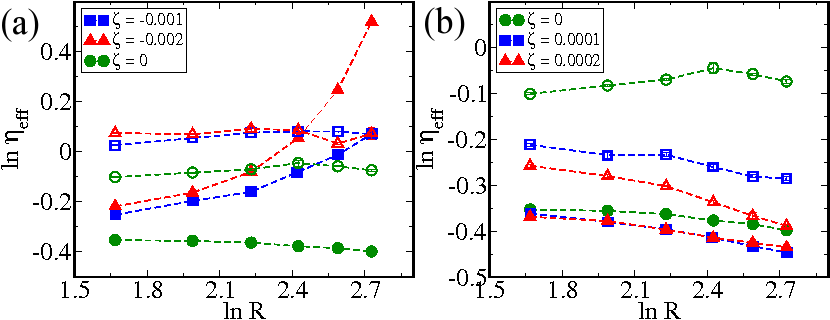}
}
\caption{Dependence of $\eta_{\rm eff}$ on $R$ for a contractile (a) and for an extensile (b) active fluid. Different symbols refer to different values of $\zeta$ (see legend). Filled and open symbols correspond to pulling along and perpendicular to the far field director respectively and dashed lines are to guide the eye. We used (periodic) cubic simulation box of volume $V=128^3$. }
\label{fig:3}       
\end{figure*}

We first analyse the case when the anchoring is planar, i.e. $\vec{n}$ lies in any direction on the plane which is locally tangential to the surface of the particle. Results are presented in Fig.~\ref{fig:3}, where the case of a passive nematic is shown as well for comparison, and where $\eta_{\rm eff}$ is plotted as a function of $R$ in a double-logarithmic scale.
In agreement with previous results and with Stokes' law~\cite{Juho_PRL}, we find that microrheology experiments performed in passive nematics ($\zeta=0$) lead to the measurements of quantitatively different viscosity, according to whether the particle is dragged along or orthogonally to the average director orientation in the bulk (with the ``orthogonal'' viscosity about a factor of 2 larger than the ``parallel'' one), in agreement with theory~\cite{Stark} and experiments~\cite{Loudet}. In active nematics, contractility increases the drag, while extensile activity reduces it -- this is in line with the macrorheology results. 

More strikingly, our results show that as soon as $\zeta\ne 0$, a completely different behaviour is observed regarding the dependence of $\eta_{\rm eff}$ on $R$: Stokes' law is no longer valid when activity is turned on. Let us focus on Fig.~\ref{fig:3} (a) first: here $\eta_{\rm eff}$ rapidly increases with $R$ when the particle is dragged along the director. 
When an extensile active fluid is considered (Fig.~\ref{fig:3} (b)) instead, $\eta_{\rm eff}$ decreases with $R$, the effect being more visible when the particle is dragged normal to the far field director.

\begin{figure*}
\resizebox{0.8\textwidth}{!}{%
  \includegraphics{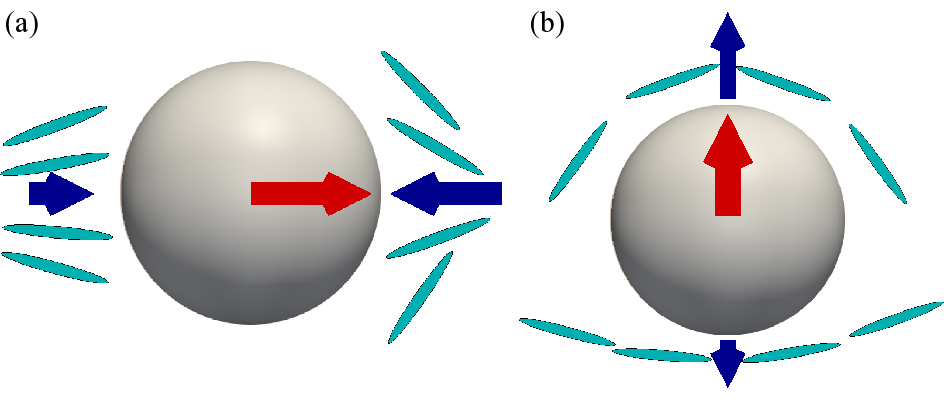}
}
\caption{Sketch of the splay and bend deformation when the colloidal particle is dragged along different directions. Panels (a) and (b) refer to dragging in a contractile and extensile fluid respectively. Blue arrows refer to the active forces, while the red one represents the external force. The sketch is further discussed in the text.}
\label{fig:4}       
\end{figure*}

We qualitatively explain these results with the following heuristic argument, sketched in Fig.~\ref{fig:4} (and summarised previously in~\cite{PRL}). If the particle were not moving, the deformations in the director field due to the anchoring condition would be perfectly symmetric, and forces on the probe due to the active stress would cancel. On the other hand, we expect that, when the particle is dragged through the active nematic fluid some asymmetries will show up in the director profile. In particular, in the case of planar anchoring, when the particle is dragged along the director, the splay in front of the particle should be greater than the one at the back (see Fig.~\ref{fig:4}a). This generates a net force that opposes the external one, if the fluid is contractile, and that favours it, in the extensile case. That the effect is more important in contractile fluids can probably be ascribed to the fact that they tend to splay more easily than their passive counterparts, while extensile ones are more stable in resisting splay~\cite{Ramaswamy_stability}. An $R$ dependence may then be expected as the splayed region depends on probe size.
On the other hand, the director field bends, both down and upstream of the particle, when the external force is applied perpendicular to the far field director (see Fig.~\ref{fig:4}b). When pulling upwards along that direction, again one expects the bending deformation on the top to be larger than the one at the bottom of the particle. These contributions combine to give a force opposing motion in contractile fluids and favouring it in extensile ones. 
Similarly to the case sketched in Fig.~\ref{fig:4}a, the effect is now larger for extensile fluids, as they tend to bend more easily than their passive counterpart, while contractile nematics are more resistant to bending~\cite{Ramaswamy_stability}.

This argument explains qualitatively our results on planar anchoring in Fig.~\ref{fig:3}. It further suggests that the orientation of the director field at the particle surface should play a key role in determining the microrheological drag of the particles. We therefore now investigate this aspect more in depth. We present in what follows both the cases of normal and of no imposed anchoring ($W=0$). Although the latter case may be unlikely realised in experiments, where the colloidal probe is unlikely to not affect the orientation of the active nematogens nearby, this calculation leads to a useful limiting case.

\begin{figure*}
\resizebox{0.9\textwidth}{!}{%
  \includegraphics{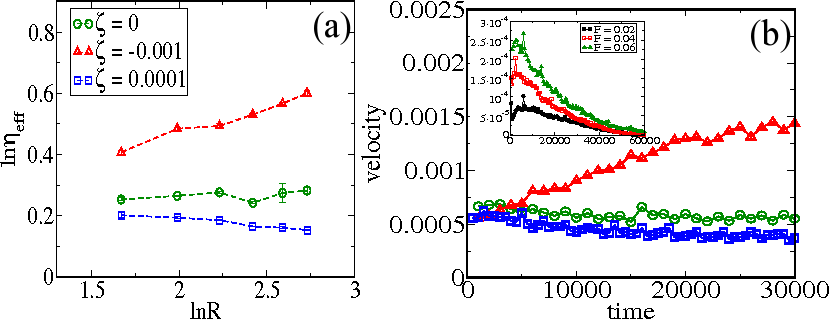}
}
\caption{(a) Dependence of the measured effective viscosity on radius for normal anchoring of the director field on the colloidal probe surface, at various $\zeta$ (see legend). The dragging direction is along the far field director. Other parameters are as in Fig.~\ref{fig:3}. (b) Velocity versus time for $F=0.12$, $R=9.3$ in the case of dragging along the far field director. The symbols are chosen according to the legend shown in (a). The inset in (b) shows the velocity of a particle of radius $R=9.3$ pulled through an extensile gel ($\zeta=0.0001$) as a function of time for three different values of the extenal force $F$. Note the presence of a non-zero `yield force' further discussed in the text (all the curves shown in the inset correspond to forces before the yield force, so that $v \to 0$ at late times). As in the previous cases, we used (periodic) cubic simulation box of volume $V=128^3$.}
\label{fig:5}       
\end{figure*}

\begin{figure*}
\centerline{
\resizebox{0.4\textwidth}{!}{%
  \includegraphics{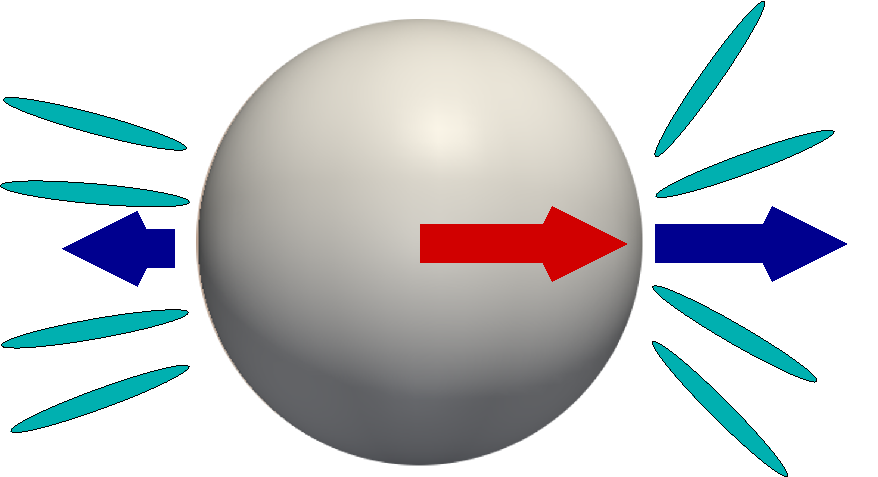}
}}
\caption{Sketch of the director field deformation for normal anchoring, and
for a probe pulled along the far field director in a contractile fluid. As before, blue arrows refer to the active forces, while the red one represents the external force.}
\label{fig:6}       
\end{figure*}

Results for normal anchoring are presented in Fig.~\ref{fig:5}. While pulling orthogonally to the far field director (see Fig.~\ref{fig:5}a) leads to results which are broadly similar to the tangential anchoring case previously shown in Fig.~\ref{fig:3}, dragging along the director leads to a very different scenario. Now, quite strikingly, the particle moves much faster in contractile than it does in extensile nematics, as is apparent from Fig.~\ref{fig:5}b. Remarkably, this can once more be explained by analysing the director deformation due to the anchoring condition, which is sketched in Fig.~\ref{fig:6}. In this case the splay is reversed with respect to the case of planar anchoring and one would expect to see an opposite effect, with contractile stresses pulling the particle forward, as we indeed observe.

\begin{figure*}
\resizebox{0.9\textwidth}{!}{%
  \includegraphics{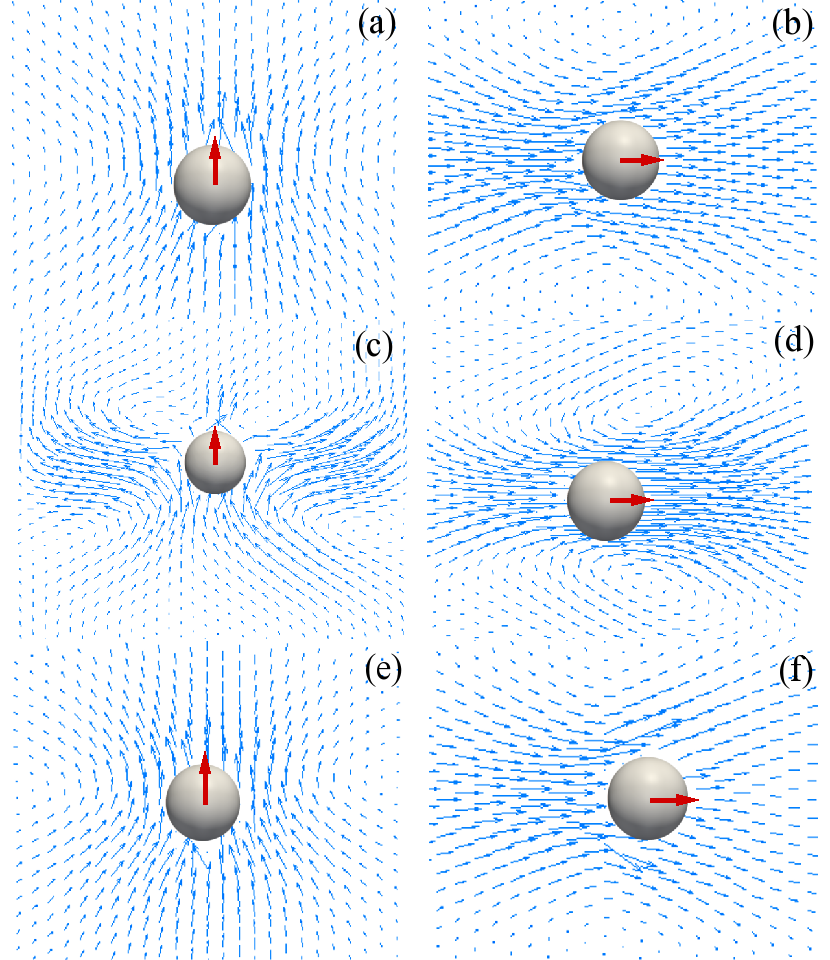}
}
\caption{Flow field patterns close to a moving particle of radius $R=11.3$ subject to a force $F=0.12$, in a passive nematic liquid crystal (a,b), in a contractile active fluid (c,d), and in an extensile active fluid (e,f). In all cases there is normal anchoring of the director field at the particle surface. In (a,c,e) the pulling direction is perpendicular to the far field director, whereas  
in (b,d,f) the colloid is dragged paralled to the far field.}
\label{fig:7}       
\end{figure*}

The scenario suggested by the cartoon in Fig.~\ref{fig:6} is fully confirmed by a detailed analysis of the flow profiles (shown in Fig.~\ref{fig:7}), around a probe particle being dragged through either passive nematic, a contractile active fluid, or an extensile one, and either along or perpendicular to the far field director. In particular, comparing panels (b), (d) and (f) one may easily note that in the contractile case the active stress leads to a flow field upstream of the particle which pulls it forward, hence facilitating its motion. In the extensile case the flow is smaller than in the passive case, again in agreement with our argument based on the director deformation.

The result for normal anchoring, that the resistance to motion may now decrease in a contractile active fluid, is particularly important. Indeed it clearly shows that not only does the viscosity inferred from microrheological experiments strongly depend on the anchoring condition, but also that the very decrease/increase of the resistance to motion is not an inherent property of the underlying fluid. This is because the estimate of $\eta_{\rm eff}$ is also affected in a nontrivial way by the anchoring conditions, which will depend ultimately on the microscopic details of the interaction between the colloidal probe and the particles that make up the active fluid.

 Intriguingly, unlike the other cases considered thus far, when pulling along the director in an extensile active fluid we find that there is a limiting force which needs to be overcome before the particle moves in steady state (see the inset in Fig.~\ref{fig:5}b, where all curves refer to forces below this limiting value, so that $v\to 0$ in steady state). This `yield force' may be seen as a microrheological analog of the yield stress in bulk rheology experiments.
 Its presence is equivalent to the absence of a linear regime which is why we do not report estimates for $\eta_{\rm eff}$ for this pulling mode.

Finally, we turn to the case of no anchoring, $W=0$.  In this limit the director field is not affected by the presence of the particle, at least when the latter is quiescent. (When the particle moves the flow field will of course deform the nearby orientation of the active fluid.) This case might be thought of as more directly probing the actual properties of the fluid, as anchoring effects play no role here, and there is only a hydrodynamic coupling between probe and fluid. 

\begin{figure*}
\resizebox{0.9\textwidth}{!}{%
  \includegraphics{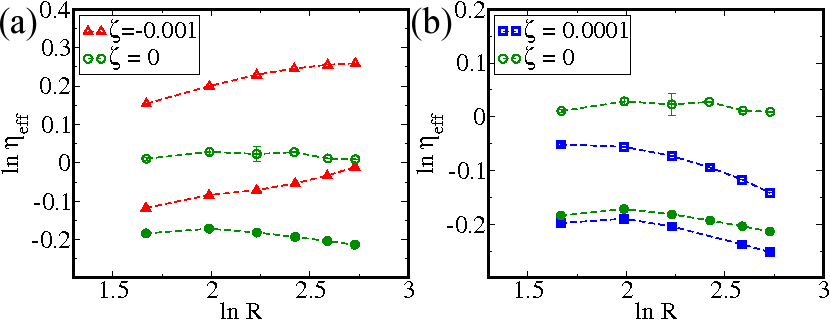}
}
\caption{Dependence of the effective viscosity on particle size for no anchoring, and different $\zeta$ (see legend). (a) and (b) describe a contractile and extensile active fluid respectively. As previously, filled and open symbols correspond to pulling along and perpendicular to the far field director respectively and we used (periodic) cubic simulation box of volume $V=128^3$. }
\label{fig:8}       
\end{figure*}

The resulting effective viscosity curves are shown in Fig.~\ref{fig:8}. They are qualitatively similar to those observed for planar anchoring (see Fig.~\ref{fig:3}): contractility determines an increase in the measured viscosity with respect to the passive case, while a decrease is observed in extensile active nematics. However, no significant $R$ dependence is found, presumably because now the director deformation is much smaller than in the $W \ne 0$ case and is less affected by the probe size.
  
All the results on microrheology presented so far were obtained with periodic boundary conditions. One may wonder whether this choice may lead to artifacts.
We therefore also consider the case in which the system is bounded by two parallel planar walls. (This provides a physical way to stabilise the quiescent phase at low activity~\cite{sriram_review,spontaneousflow_I,spontaneousflow_II,spontaneousflow_III} although periodic boundary conditions also do this.) 

\begin{figure*}
\resizebox{0.9\textwidth}{!}{%
  \includegraphics{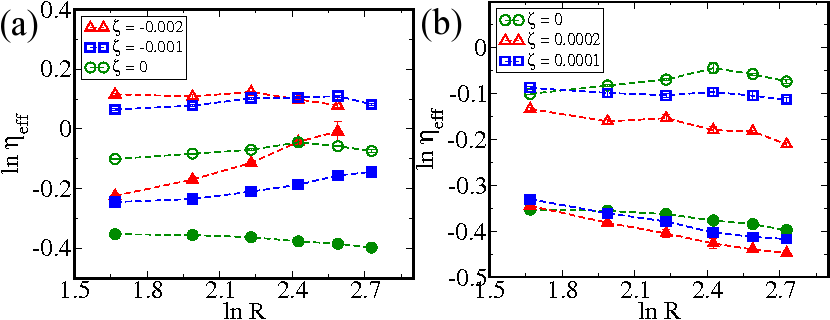}
}
\caption{Dependence of $\eta_{\rm eff}$ on $R$ for (a) contractile and (b) extensile fluids, with particle anchoring as in Fig.~\ref{fig:3}, but with strong homogeneous anchoring of the director field at the wall of the container. The anchoring determines in this case the far field director profile. All parameters are as in Fig.~\ref{fig:3}, apart from the system size which is 64 along the direction orthogonal to the walls, and 128 in the other two directions.}
\label{fig:9}       
\end{figure*}

The results in this case are presented in Fig.~\ref{fig:9}~\cite{note-walls}. Pleasingly, it can be seen that they confirm, at least qualitatively, what was previously found with periodic boundary conditions. The effect of activity is now slightly smaller, probably because the system is half the size of the one simulated for the data shown in Fig.~\ref{fig:3}, so that the director deformation close to the particles may be slightly less.

\begin{figure*}
\resizebox{0.9\textwidth}{!}{%
  \includegraphics{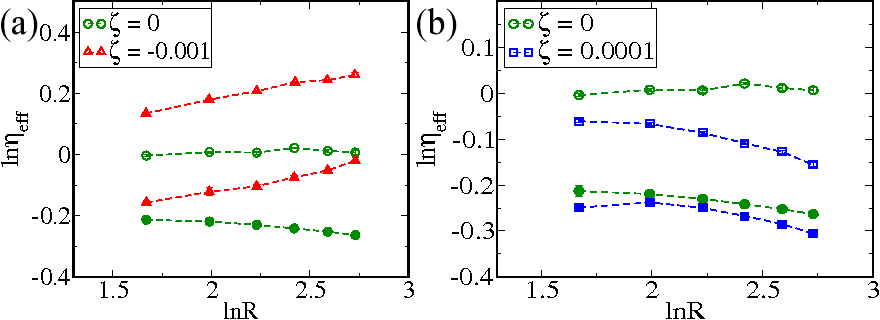}
}
\caption{Dependence of the effective viscosity on particle size with ${\mathsf Q}=0$ on its surface, and different values of $\zeta$ (see legend). (a) and (b) refer to a contractile and extensile active fluid respectively. As previously, filled and open symbols correspond to pulling along and perpendicular to the far field director respectively and we used (periodic) cubic simulation box of volume $V=128^3$.}
\label{fig:10}       
\end{figure*}

One important constraint which our model should fulfill to reliably address an experimentally realisable situation is that the probe used to obtain the microrheology data should be considerably larger than active nematogens. This is likely to be the case in practice if actomyosin (or some other molecular active fluid) is chosen as a host medium. If bacterial suspensions are considered, this condition would not easily be met, as the size of probes used in microrheology experiments is usually between 1 and 10$\mu$m~\cite{waigh}, therefore comparable to the one of bacteria ({\it E.coli} for example is about $\sim$1$\mu$m in size). If a sufficiently dense suspension is considered, the continuous approximation still holds and our model still gives a good description of the system. However, one may imagine that in this case, rather than imposing any particular anchoring for the bacterial director on its surface, the probe particle presence may simply enhance disorder in the bacterial orientation locally.
One way to model this is to impose a null value of the amplitude of order $S_0$ for ${\mathsf Q}$ in eq.~\ref{eq:fed_surf}. The resultant director field in the proximity of the particle is very similar to the $W = 0$ case, as expected.
The microrheology results are shown in Fig.~\ref{fig:10}: $\eta_{\rm eff}$ is again found to be larger with respect to the passive one in contractile nematics, and smaller in extensile ones, in agreement with most of our previous results. A rather mild dependence on $R$ is observed for pulling both parallel and orthogonally to the far-field director and, remarkably, no significant difference is observed between these two cases in terms of the $R$ dependence with this boundary condition. This is probably because the chosen boundary condition melts the active nematic close to the particle surface, so that the deformations of the orientational order close to the particles should not much depend on the dragging direction. 

\section{Discussion and conclusion}
\label{sec:3}
In conclusion, we presented numerical results on the bulk rheology and on the microrheology of active fluids, specifically active nematics, consistent with previous work~\cite{spontaneousflow_I,spontaneousflow_II,spontaneousflow_III,fieldingPRL}. 
We showed that when shearing an active nematic fluid in a quasi-one dimensional geometry, the effective viscosity strongly depends on system size, and also on the imposed anchoring of the active nematogen orientation at the wall. Our numerical results are in very good agreement with analytical estimates, obtained by linearising the relevant hydrodynamic equations of motion. 
Our results thus demonstrate that bulk rheology experiments performed on active nematics should strongly depend on the details of the measurement setup, precluding any identification of the effective viscosity of an active fluid as an inherent property of the material. 

We also simulated the simplest possible microrheology experiment in an active fluid. That is, we applied a constant force, either parallel or orthogonal to the far field director, to a spherical probe particle embedded in an active nematic, and measured the steady state velocity it attains, which leads to an estimate for the drag on the particle. If Stokes' law held in this active context, we would expect the drag coefficient to be linearly proportional to the particle size and to an effective viscosity, $\eta_{\rm eff}$. 
We presented results for both planar and normal anchoring, which show that the colloidal drag is in both cases strongly non-Stokesian, as the dependence on size is highly non-linear. We gave a qualitative explanation of this non-Stokesian drag in terms of the local deformation of the director orientation close to the particle surface.
Our argument also explains why planar and normal anchoring lead to qualitatively different results: for example, when a colloidal probe is pulled along the far field director, its drag is larger than that in a passive nematic when the anchoring is planar, but it decreases for normal anchoring.

We also presented numerical results for the effect on microrheology experiments, with different boundary conditions. In particular, we simulated the case in which the sample is bounded by solid walls, as opposed to that in which periodic boundary conditions are considered. Pleasingly, we found that our results in the two cases are similar.
Finally we also considered both an idealised case in which the colloidal probe does not affect the orientation of the active nematogens in equilibrium, and the extreme case in which instead the order parameter is fixed to zero on its surface. With both boundary conditions, we found that the results qualitatively resembled those obtained with planar anchoring, and we observed that the drag on the particle, hence $\eta_{\rm eff}$, increases in contractile fluids, and decreases in extensile ones, although the dependence on size was milder.

We hope our results will stimulate further experimental work, on both the bulk rheology and the microrheology of active fluids such as actomyosin solutions or very dense bacterial suspensions. (The latter can be probed via optical tweezers~\cite{Yao}.) 

We acknowledge support from the EU training network 
ITN-COMPLOIDS (FP7-234810), as well as from EPSRC grants
 EP/E030173/1, EP/I030298/1 and EP/I004262/1. MEC is supported by the
Royal Society.


\begin{thebibliography}{0}
\bibitem{noteactomyosin} It is important to note that in order to
experimentally test our model, which assumes uniform activity, 
one needs to avoid myosin clustering (observed e.g. in Ref.~\cite{silva}).
\bibitem{activity_oom} Note that the absolute values of activity chosen in the contractile and in the extensile case differ of an order of magnitude. This is because spontaneous bulk flow shows up at much smaller $|\zeta|$ in extensile than in contractile active fluids.
\bibitem{note-walls} 
Hydrodynamic interactions between different images of the particle due to periodic boundary conditions are now still there in the two directions parallel to the wall, however they are no longer described by the theory of~\cite{Hasimoto}. We therefore corrected our data through the deviation of the simple fluid viscosity measured in the same conditions, from its theoretical value, which corresponds to an infinite box size parallel to the boundaries.
\end{thebibliography}
\end{document}